\documentclass[12pt]{article}
\usepackage{cite}

\begin{document}
\title{A New Proof of Existence of a Bound State in the Quantum Coulomb
Field}
\author{Andrzej Staruszkiewicz\\
Marian Smoluchowski Institute of Physics,
Jagellonian University\\
Reymonta 4, 30-059 Krak\'ow, Poland\\
e-mail: {\tt astar@th.if.uj.edu.pl}
}
\maketitle

\begin{abstract}
Let $S(x)$ be a massless scalar quantum field which lives on the
three-dimensional hyperboloid $xx= (x^0)^2-(x^1)^2-(x^2)^2-(x^3)^2=-1.$
The classical action is assumed to be
$(\hbar=1=c)(8\pi e^2)^{-1}\int dx g^{ik}\partial_i S\partial_k S$, where
$e^2$ is the coupling constant, $dx$ is the invariant measure on the de
Sitter hyperboloid $xx=-1$ and $g_{ik}, i,k=1,2,3$, is the internal
metric on this hyperboloid. Let $u$ be a fixed four-velocity i.e. a fixed
unit time-like vector. The field $S(u)=(1/4 \pi)\int dx\delta(ux)S(x)$is
smooth enough to be exponentiated, being an average of the operator
valued distribution $S(x)$ over the entire Cauchy surface $ux=0$.
We prove that if $0<e^2<\pi$, then the state $|u\rangle=\exp(-iS(u))\mid
0\rangle$, where $\mid 0\rangle$ is the Lorentz invariant vacuum state,
contains a normalizable eigenstate of the Casimir operator
$C_1=-(1/2)M_{\mu\nu}M^{\mu\nu}$; $M_{\mu\nu}$ are generators of
the proper orthochronous Lorentz group. The eigenvalue is
$(e^2/\pi)(2-(e^2/\pi)).$ This theorem was first proven by the Author in
1992 in his contribution to the Czy¿ Festschrift, see Erratum {\it Acta Phys.
Pol. B} {\bf 23}, 959 (1992). In this paper a completely different proof
is given: we derive the partial, differential equation satisfied by the
matrix element $\langle u\mid \exp (-\sigma C_1)\mid u\rangle, \sigma >
0$, and show that the function $\exp (z)\cdot (1-z)\cdot \exp[-\sigma z
(2-z)], z= e^2/ \pi$, is an exact solution of this differential equation,
recovering thus both the eigenvalue and the probability of occurrence of
the bound state. A beautiful integral is calculated as a byproduct.

\end{abstract}
\vspace{-0.3cm}
\vspace{-0.3cm}
\section{Introduction}
We use mechanical units such that $\hbar=1=c$. We use electric units
such that the fine structure constant is equal to $1/e^2$, where $e$ is
the electron's charge. We use space-time metric such that $g(x,x)=xx=
(x^0)^2-(x^1)^2-(x^2)^2-(x^3)^2$ is the square of the length of the
vector $x$.

In Ref.~[1] we were led to consider the following theoretical structure.
Let $x^{\mu}, \mu=0,1,2,3$, denote space-time Cartesian coordinates in
an inertial reference frame. The equation $xx=-1$ defines a subspace
which is locally a three-dimensional space-time and is maximally
symmetric, admitting six Killing vectors. Thus it is a three-dimensional
analogue of  de Sitter space-time and will be called simply
three-dimensional de Sitter space-time. A scalar massless quantum field
is assumed to live on the de Sitter space-time $xx=-1.$ Its classical
action is assumed to be 
\begin{equation}
\frac{1}{8\pi e^2}\int\limits_{xx=-1}dx g^{ik}\partial_i S \partial _k
S\,,
\end{equation}
where $e^2$ is the coupling constant, $dx$ is the invariant measure on
the de Sitter hyperboloid $xx=-1$ and $g_{ik}, i,k=1,2,3,$ is the
internal metric on this hyperboloid. The above action has the following
symmetries: the Lorentz symmetry, which, via the first Noether theorem,
gives rise to six constants of motion $M_{\mu\nu}=-M_{\nu\mu}$ and the
``gauge'' symmetry $S(x)\to S(x)+\,$const, which, again via the first
Noether theorem, gives rise to the additional constant of motion called
the total charge,
\begin{equation}
Q={-1\over 4\pi e}\int\limits_{\rm C.S.} d{\mit\Sigma}^i\partial_i S\,.
\end{equation}
Here C.S. means a Cauchy surface in the de Sitter hyperboloid $xx=-1$
and $d{\mit\Sigma}^i$ is the integration element on this surface.

A quantum field theory is obtained if we assume that
\begin{equation}
[M_{\mu\nu},\, S(x)]={1\over i} \left(
x_\mu{\partial\over\partial x^\nu}-x_\nu{\partial\over\partial x^\mu}\right)S(x)
\end{equation}
and that there exists a state $|0\rangle$ such that
\begin{equation}
M_{\mu\nu}|0\rangle=0\,,\quad \langle 0| M_{\mu\nu}=0\,,\quad \langle
0|0\rangle =1\,.
\end{equation}
Eqs (3) and (4) can hold only if
\begin{equation}
[Q,S(x)]=ie
\end{equation}
and
\begin{equation}
Q|0\rangle =0\,,\quad \langle0|Q=0\,,\quad \langle 0|0\rangle =1\,.
\end{equation}
There are many misleading or erroneous statements in the literature on
the vacuum state in de Sitter space-time; for this reason the reader is
invited to check the consistency of Eqs (3)--(5), and (6) with the help
of Ref.~[1].

The quantum field $S(x)$ is an operator valued distribution and cannot
be a subject of nonlinear operations. It is, however, a very fortunate
circumstance that Cauchy surfaces in the de Sitter space-time $xx=-1$
are compact. For this reason averaging over a Cauchy surface has the
quality of smearing out with an arbitrarily smooth function of compact
support in  QED.

Let us choose a fixed unit time-like vector $u$. The quantum field
\begin{equation}
S(u)={1\over 4\pi}\int\limits_{xx=-1} dx \delta(ux)S(x)
\end{equation}
is the average of the field $S(x)$ over the compact section of the
space-like plane $ux=0$ and the de Sitter hyperboloid $xx=-1$ and is
smooth enough to be exponentiated. $S(u)$ is a quantum field which lives
in the Lobachevsky space of four-velocities $uu=+1$. It is easy to see
that
\begin{equation}
[M_{\mu\nu},\,S(u)] = {1\over i}\left(
u_\mu{\partial\over\partial u^\nu}-u_\nu{\partial\over\partial u^\mu}\right)S(u)
\end{equation}
and
\begin{equation}
[Q,\,S(u)]=ie\,.
\end{equation}
Using the smooth quantum field $S(u)$ we can consistently form a charged
state
\begin{equation}
|u\rangle =\exp (-iS(u))|0\rangle
\end{equation}
which is an eigenstate of the total charge $Q$:
\begin{equation}
Q|u\rangle =e|u\rangle\,,\quad \langle u|u\rangle=1\,.
\end{equation}

We shall investigate in this paper some properties of charged states of
the form $\exp (-iS(u))|0\rangle$. In particular, we shall give a
completely new proof of the theorem that the spectral contents of the
state $\exp(-iS(u))|0\rangle$ in the regions $0<e^2<\pi$ and $e^2>\pi$
are different. By spectral content we mean the way in which a given
state can be represented as a superposition of eigenstates of the first
Casimir operator
\begin{equation}
C_1=-\textstyle{1\over2} M_{\mu\nu} M^{\mu\nu}\,.
\end{equation}

\section{Calculation of the matrix element\\ $\langle u|\exp(-\sigma
C_1)|u\rangle\,,\ \sigma>0$}

To save space we shall write $\nabla_{\mu\nu}(u)$ instead of
$$
u_\mu{\partial\over\partial u^\nu}-u_\nu{\partial\over \partial u^\mu}\,.
$$
In expressions like $\nabla_{\mu\nu} (u) S (u)$ one can even drop the
first argument $u$ because the argument $u$ of $S(u)$ indicates the
variable with respect to which the differentiation is carried out. Let
us note first that
\begin{equation}
[S(u)\,,\ S(v)]=0
\end{equation}
for each pair of points $u,v$ in the Lobachevsky space $uu=+1$. This is
so because the definition (7) of $S(u)$ picks up only the even part of
the field $S(x)$ i.e. the part $(1/2)[S(x)+S(-x)]$ and even parts do
commute with each other on the strength of canonical commutation
relations (3). As an obvious consequence we have that
\begin{equation}
[S(u)\,,\ \nabla_{\mu\nu}S(u)]=0\,.
\end{equation}
Consider now $[S(u)\,,\, C_1]$. We have
\begin{eqnarray}
[S(u)\,,\, C_1]&=&{1\over 2}[M_{\mu\nu} M^{\mu\nu}\,, \, S(u)]\nonumber\\
&=& {1\over 2}\left\{ {1\over i}\nabla_{\mu\nu} S(u)\cdot M^{\mu\nu}+
M^{\mu\nu}\cdot {1\over i}\nabla_{\mu\nu} S(u)\right\}\,.
\end{eqnarray}
Therefore
\begin{equation}
[S(u)\,,\ [S(u)\,,\, C_1]]=\nabla_{\mu\nu} S(u)\nabla^{\mu\nu} S(u)
\end{equation}
and
\begin{equation}
[S(u)\,,\,[S(u)\,,\,[S(u)\,,\,C_1]]]=0\,.
\end{equation}
Consider now the matrix element $\langle v|\exp(-\sigma C_1)|u\rangle$,
where $v$ is a fixed four-velocity different from $u$ and $\sigma>0$.
Using Eqs (14)--(16) and (17) in an obvious way we have
\begin{equation}
-{\partial\over\partial\sigma}\langle v|\,e^{-\sigma
C_1}|u\rangle=\langle v|\,e^{-\sigma C_1}\,e^{-iS(u)}{-1\over
2}\nabla_{\mu\nu}S(u)\nabla^{\mu\nu}S(u)|0\rangle\,.
\end{equation}
On the other hand, let us apply the Laplace--Beltrami operator
${\mit\Delta}(u)=-(1/2)\nabla^{\mu\nu}(u)\nabla_{\mu\nu}(u)$ to the
matrix element $\langle v|\exp(-\sigma C_1)|u\rangle$. Taking into
account that ${\mit\Delta}(u)S(u)=0$ as a consequence of the equation of
motion ${\mit\Delta}(x)S(x)=0$ we have
\begin{equation}
{\mit\Delta}(u)\langle v|\,e^{-\sigma C_1}|u\rangle=\langle v|\,e^{-\sigma
C_1}\,e^{-iS(u)}{-1\over
2}\nabla_{\mu\nu}S(u)\nabla^{\mu\nu}S(u)|0\rangle\,.
\end{equation}
Comparing (18) and (19) we see that
\begin{equation}
\left\{{\partial\over\partial\sigma}-\mit\Delta(u)\right\}\langle
v|\,e^{-\sigma C_1}|u\rangle=0\,.
\end{equation}
This means that the matrix element $\langle v|\exp(-\sigma
C_1)|u\rangle$
is a solution of the heat transport equation in the Lobachevsky space of
four-velocities $uu=+1$. The initial value for this solution is
obviously the matrix element $\langle v|u\rangle$ which was calculated
in Ref.~[1] as $\exp(-(e^2/\pi)(\lambda\coth\lambda-1))$, where $\lambda$
is the hyperbolic angle between $u$ and $v$:

\begin{equation}
g_{\mu\nu} u^\mu v^\nu =\cosh \lambda\,.
\end{equation}
To solve the Cauchy problem for the heat transport equation (20) we
apply the standard procedure: we represent the initial value as a
superposition of plane waves, solve the heat transport equation for each
plane wave, and represent the final value as a superposition of time
evolved plane waves. However, since we are in the Lobachevsky space of
four-velocities $uu=+1$, we have to apply plane waves in Lobachevsky
space which Gelfand, Graev, and Vilenkin described in their great book
[2]. This means that we have to apply Eqs (20) and (21) which Gelfand,
Graev, and Vilenkin give on page 477 of their book. These equations give
the Fourier transform and its inverse in Lobachevsky space. The result
of this obvious procedure is summarized in the following lemma: suppose
that $f(u;0)$ is the initial value for the function $f(u;\sigma)$ which
solves the heat transport equation in the Lobachevsky space $uu=+1$,

$$
\left\{{\partial\over\partial\sigma}-{\mit\Delta}(u)\right\}
f(u;\sigma)=0\,.
$$
Then
\begin{equation}
f(u;\sigma)={1\over 2\pi^2}\int du'
f(u';0){1\over\sinh\lambda}\int\limits^\infty_0 d\nu \nu\,e^{-\sigma
(1+\nu^2)}\sin(\nu\lambda)\,,
\end{equation}
where $du$ is the invariant measure in the Lobachevsky space $uu=+1$ and
$\lambda$ is the hyperbolic angle between the observation point $u$ and
the integration point $u'$. The second integral in (22) can be
calculated. In this way we obtain
\begin{equation}
f(u;\sigma)={e^{-\sigma}\over (4\pi\sigma)^{3/2}}\int du' f
(u';0){\lambda\over \sinh\lambda}\,e^{-{\lambda^2\over 4\sigma}}\,,
\end{equation}
where $\lambda$ is the hyperbolic angle between the observation point
$u$ and the integration point $u'$.

Now, we took the matrix element $\langle v|\exp(-\sigma C_1)|u\rangle$
in order to be able to differentiate with respect to $u$ while leaving $v$
untouched. In fact, however, we are interested in the matrix element
$\langle u|\exp(-\sigma C_1)|u\rangle$. Geometrically this means that we
have to take in Eq.~(23) the observation point $u$ at the origin of the
spherically symmetric distribution $f(u';0)$. Introducing the spherical
coordinates
\begin{eqnarray}
u'\,^{0}&=&\cosh\psi\,,\nonumber\\
u'\,^{1}&=&\sinh\psi\sin\vartheta\cos\varphi\,,\nonumber\\
u'\,^{2}&=&\sinh\psi\sin\vartheta\sin\varphi\,,\nonumber\\
u'\,^{3}&=&\sinh\psi\cos\vartheta\,,
\end{eqnarray}
and taking the solution at the origin of spherical symmetry we have
finally $(z=e^2/\pi)$
\begin{equation}
\langle u|\,e^{-\sigma C_1}|u\rangle={1\over
2\sqrt\pi}{e^{-\sigma}\over \sigma^{3/2}}\int\limits^\infty_0
d\psi\sinh \psi\,e^{-z(\psi\coth\psi-1)}\cdot\psi\,e^{-{\psi^2\over
4\sigma}}\,.
\end{equation}

\section{The differential equation satisfied by the matrix element
$\langle u|\exp(-\sigma C_1)|u\rangle$}

It is remarkable that the matrix element $\langle u|\exp(-\sigma
C_1)|u\rangle$ satisfies a certain partial differential equation. This
equation, as well as the equation arrived at later on, satisfied by the
resolvent $\langle u|(C_1-\lambda)^{-1}|u\rangle$, is obviously a trace
of some deeper structure which, for the time being, we fail to
understand.

In fact let us put
\begin{equation}
f(\nu, z)=\int\limits^\infty_0 d\psi
\sinh\psi\,e^{-z\psi\coth\psi}\cdot \psi\,e^{-\nu\psi^2}\,,\quad
\nu>0\,.
\end{equation}
Then
\begin{eqnarray}
{\partial f\over\partial z}&=& -\int\limits^\infty_0
d\psi\,e^{-z\psi\coth\psi}\cosh\psi\cdot
\psi^2\,e^{-\nu\psi^2}\nonumber\\
&=&\int\limits^\infty_0 d\psi\sinh\psi{d\over d\psi}\left\{
e^{-z\psi\coth\psi}\cdot\psi^2\,e^{-\nu\psi^2}\right\}\nonumber\\
&=& z{\partial f\over\partial z}+z{\partial^2 f\over\partial
z^2}+z{\partial f\over\partial \nu}+2f+2\nu{\partial f\over\partial \nu}
\end{eqnarray}
which means that the function $f(\nu,z)$ is a solution of the partial
differential equation
\begin{equation}
(z-1){\partial f\over\partial z}+ z{\partial^2 f\over\partial
z^2}+z{\partial f\over \partial\nu}+2\nu{\partial
f\over\partial\nu}+2f=0\,.
\end{equation}
Having this equation we can multiply the function
$f(\nu,z)=f(1/4\sigma,z)$ by trivial factors
$\sigma^{-3/2}\exp(z-\sigma)$ and obtain the following lemma: let
$\langle u|\exp(-\sigma C_1)|u\rangle=c(\sigma,z)$;
then the function $c(\sigma,z)$ satisfies the partial differential
equation
\begin{equation}
z{\partial^2 c\over \partial z^2}-(z+1){\partial c\over\partial
z}-2\sigma(1+2\sigma z){\partial
c\over\partial\sigma}-2\sigma(1+3z+2\sigma z)c=0\,.
\end{equation}

\section{The eigenvalue and the probability of occurrence of the bound
state of the Casimir operator $C_1$ in the state
$|u\rangle=\exp(-iS(u))|0\rangle$}

The Casimir operator $C_1$ is known to have no lower bound; its
eigenvalues in the so called main series of irreducible unitary
representations of the proper orthochronous Lorentz group are [2]
$1+\nu^2-n^2$, where $\nu$ is a real number and $n$ is an integer.
However, $n$ is proportional to the eigenvalue of the second Casimir
operator
\begin{equation}
C_2=M_{01}M_{23}+M_{02}M_{31}+M_{03}M_{12}
\end{equation}
which obviously annihilates all spherically symmetric states.
$|u\rangle=$\break$\exp(-iS(u))|0\rangle$ is spherically symmetric in the rest
frame of $u$. Therefore in the subspace of spherically symmetric states
the Casimir operator $C_1$ does have a lower bound. This means that the
asymptotic behaviour of solutions of Eq.~(29) for $\sigma\to\infty$ is
determined by the state with the lowest eigenvalue.

Assume that the spectral decomposition of $C_1$ does contain a bound
state with the lowest eigenvalue. Then Eq.~(29) has to have for
$\sigma\to\infty$ an asymptotic solution of the form
\begin{equation}
c_0(\sigma,z)=A(z)\,e^{-\sigma B(z)}\,.
\end{equation}
Putting this into Eq.~(29) we obtain on the left hand side a polynomial
of second degree in $\sigma$ with coefficients depending on $z$. Thus
for $\sigma\to\infty$ all three coefficients have to vanish. This gives
us three ordinary differential equations for two functions $A(z)$ and
$B(z)$. Remarkably enough, all three equations can be simultaneously
solved with the result
\begin{equation}
A(z)=(1-z)\,e^z\,,\quad B(z)=z(2-z)\,.
\end{equation}
In this way we have the following lemma: the function
$c_0(\sigma,z)=(1-z)\,e^z\cdot\exp[-\sigma z(2-z)]$ is an exact
solution of the partial differential equation~(29). This obviously
suggests that in the state $|u\rangle$ there is a bound state of the
Casimir operator $C_1$ with the eigenvalue $z(2-z)$ and the probability
of occurrence $(1-z)\,e^z$. This probability cannot be negative which
means that the state can exist only for $0<z<1$. It is thus seen that
the coupling constant $z=e^2/\pi=1$ is critical and separates two
kinematically different regimes of the theory. For $0<z<1\,,\
0<z(2-z)<1$ which means that the bound state belongs to the so called
supplementary series of irreducible unitary representations of the
proper orthochronous Lorentz group [2], since for the main series
$1<C_1<\infty$ [2].

\section{Calculation of the resolvent $\langle
u|(C_1-\lambda)^{-1}|u\rangle$}

Let us multiply Eq.~(25) by $\exp(\lambda\sigma)$, assume that $\lambda$
is smaller than the smallest eigenvalue of the Casimir operator $C_1$
present in the spectral decomposition of the matrix element $\langle
u|\exp(-\sigma C_1)|u\rangle$ and integrate both sides over $\sigma$,
$0<\sigma<\infty$. All integrals on the right hand side are absolutely
convergent and their order can be interchanged. In this way we obtain
\begin{eqnarray}
&&\left\langle u\left\vert{1\over C_1-\lambda}\right\vert u\right\rangle\nonumber\\
&&={1\over 2\sqrt\pi}\int\limits^\infty_0
d\psi\sinh\psi\,e^{-z(\psi\coth\psi-1)}\cdot\psi\int\limits^\infty_0 d\sigma
\sigma^{-3/2}\,e^{-\sigma(1-\lambda)-{\psi^2\over 4\sigma}}\nonumber\\
&&=\int\limits^\infty_0
d\psi\sinh\psi\,e^{-z(\psi\coth\psi-1)-\psi\sqrt{1-\lambda}}\,.
\end{eqnarray}
The last integral exists for $1-z-\sqrt{1-\lambda}<0$ i.e. for
$\lambda<z(2-z)$ which was assumed from the very beginning.

It is again remarkable that the last integral which is not given in the
Ryzhik and Gradshteyn Tables (I checked it in the VI$^{\rm th}$ American
Edition) can be calculated exactly with the help of the partial
differential equation which this integral is a solution of.

In fact, consider the integral
\begin{equation}
F(x,y)=\int\limits^\infty_0 d\zeta e^{-x\zeta-y\zeta\coth\zeta}
\end{equation}
which exists for $x+y>0$. Differentiating and integrating by parts, as
in Section 3, one can show that this integral fulfills the partial
differential equation
\begin{equation}
F+x{\partial F\over \partial x}+y {\partial F\over \partial
y}+y\left({\partial^2 F\over \partial y^2}-{\partial^2 F\over\partial
x^2}\right)=0\,.
\end{equation}
This is a hyperbolic equation for which the straight line $x+y=0$ is the
boundary of the domain of influence of the positive $x$ axis $y=0$,
$x>0$. Hence we can try to solve the Cauchy problem with the initial
data on the positive $x$ axis $y=0$, $x>0$. We have that
\begin{equation}
F(x,0)=\int\limits^\infty_0 d\zeta\,e^{-x\zeta}={1\over x}\,.
\end{equation}
This singularity must propagate to the left since $F(x,y)$ is regular
for\break $x+y>0$. The function $1/(x+y)$ is an exact solution of Eq.~(35).
Therefore $F(x,y)=1/(x+y)$ plus a solution of Eq.~(35) which vanishes
for $y=0$. We have
\begin{equation}
{\partial F\over\partial y}\Bigg\vert_{y=0}=\int\limits^\infty_0
d\zeta\,e^{-x\zeta}(-)\zeta\coth\zeta\,.
\end{equation}
Subtracting from this function the function
\begin{equation}
{\partial\over \partial y}{1\over(x+y)}\Bigg\vert_{y=0}=-{1\over x^2}
\end{equation}
I can say that $F(x,y)=1/(x+y)$ plus a solution of Eq.~(35) which
vanishes at $y=0$ and whose $y$ derivative at $y=0$ is equal to
\begin{equation}
{1\over x^2}-\int\limits^\infty_0 d\zeta
\zeta\,e^{-x\zeta}\coth\zeta=-2\sum^\infty_{n=0}{1\over (x+2n+2)^2}\,.
\end{equation}
>From the superposition principle it is seen that the problem is thus
reduced to the following one: find a solution of Eq.~(35) which vanishes
for $y=0$ and whose $y$ derivative at $y=0$ is equal to $-2/(x+2n+2)^2$.
One can check that this solution is equal to
\begin{equation}
-2y{(x+2n+2-y)^n\over (x+2n+2+y)^{n+2}}\,.
\end{equation}
Therefore for $x+y>0$
\begin{equation}
\int\limits^\infty_0 d\zeta\,e^{-x\zeta-y\zeta\coth\zeta}={1\over
x+y}-2y\sum^\infty_{n=0}{(x+2n+2-y)^n\over (x+2n+2+y)^{n+2}}
\end{equation}
which is a nice result not to be found in the Ryzhik and Gradshteyn
Tables.

The result (41) allows us to calculate the resolvent (33) since
$\sinh\psi=(1/2)(\exp\psi-\exp(-\psi))$ and the integral (33) is seen to
be of the form (34). Making the obvious substitutions we obtain for
$0<z<1$:
\begin{eqnarray}
\left\langle u\left\vert{1\over C_1-\lambda}\right\vert u\right\rangle&=&
{(1-z)e^z\over z(2-z)-\lambda}\nonumber\\
&&+2 z^2 e^z\sum^\infty_{n=0}{(\sqrt{1-\lambda}+2n+1-z)^{n-1}\over
(\sqrt{1-\lambda}+2n+1+z)^{n+2}}\,.
\end{eqnarray}
This formula shows at once the eigenvalue of the bound state, the
probability of its occurrence and the cut $1\leq\lambda<\infty$ which
reflects the contribution from the main series of irreducible unitary
representations of the proper orthochronous Lorentz group.

We see from the formula (42) that the bound state cannot exist for $z>1$
since the probability of occurrence cannot be negative. This can also be
seen from the calculation leading to the formula (42). For $x+y>1$ we
have instead of (41)
\begin{eqnarray}
&&\int\limits^\infty_0
d\zeta\sinh\zeta\,e^{-x\zeta-y\zeta\coth\zeta}\nonumber\\
&&={1/2\over x+y-1} - {1/2\over x-y+1}+2y^2\sum^\infty_{n=0}
{(x+2n+1-y)^{n-1}\over(x+2n+1+y)^{n+2}}\,.
\end{eqnarray}
This formula can be derived in the same way as the previous one, given
in Eq.~(41).
Making the obvious substitutions we obtain for $z>1$:
\begin{eqnarray}
\left\langle u\left\vert{1\over C_1-\lambda}\right\vert u\right\rangle &=&
e^z\Bigg\{{1\over 2(\sqrt{1-\lambda}+z-1)}-{\sqrt{1-\lambda}+3z+1\over
2(\sqrt{1-\lambda}+z+1)^2}\nonumber\\
&&+2z^2\sum^\infty_{n=1}{(\sqrt{1-\lambda}+2n+1-z)^{n-1}\over
(\sqrt{1-\lambda}+2n+1+z)^{n+2}}
\Bigg\}\,.
\end{eqnarray}
This resolvent has only the cut $1\leq\lambda<\infty$ which reflects
contribution from the main series.

\section{A method to calculate the averages $\langle u|(C_1)^n|u\rangle$
for integer $n$}

Differentiating both sides of Eq.~(42) with respect to $\lambda$ and
putting $\lambda =0$ we can calculate all the averages of the form
$\langle u|(C_1)^{-n}|u\rangle$, $n=1,\,2,\,3,\dots\,.$ On the other
hand there is no simple way to calculate the averages $\langle
u|(C_1)^n|u\rangle$, $n=1,\,2,\,3,\dots\,.$ Professor Wosiek and dr 
Rostworowski calculated from first principles the following averages
$(z=e^2/\pi)$:
\begin{eqnarray}
\langle u|C_1|u\rangle&=&2z\nonumber\\
\langle u|(C_1)^2|u\rangle&=&{20\over 3} z^2\nonumber\\
\langle u|(C_1)^3|u\rangle&=&{8\over 9}z^2(12+35 z)\nonumber\\
\langle u|(C_1)^4|u\rangle&=& {16\over 45} z^2(192+560 z+525
z^2)\nonumber\\
\langle u|(C_1)^5|u\rangle&=&{32\over 9} z^2(192+704 z+840 z^2+385
z^3)\nonumber\\
\langle u|(C_1)^6|u\rangle&=& {64\over 945} z^2(147456+647808 z +977760
z^2+ 646800 z^3\nonumber\\ &&+175175 z^4)\nonumber\\
&&\hspace{-2.5
cm}..............................\,.
\end{eqnarray}
One can see from these expressions that these averages increase so
quickly that the autocorrelation function $\langle u|\exp(i\sigma
C_1)|u\rangle$ cannot have a convergent Taylor series in $\sigma$. This
is not a problem, of course. Autocorrelation functions do not have to be
analytic at the origin. Nevertheless, we have observed the following
``experimental'' fact: the averages (45) can be recovered from the
differential equation (29) in the following way.

We write formally
\begin{equation}
\langle u|e^{-\sigma C_1}| u\rangle=c(\sigma,z)=\sum^\infty_{n=0}
{(-1)^n\over n!} \sigma^n c_n(z)\,;
\end{equation}
we put this into Eq.~(29) and obtain the recurrence relation
\begin{equation}
zc''_n-(z+1) c'_n-2n c_n
= 4n(n-1)zc_{n-2} -n[4z(n-1)+2(1+3z)]c_{n-1}\,.
\end{equation}
Knowing that $c_0(z)=1$, $c_1(z)=2z$ and assuming that $c_n(z)$ is a
polynomial of degree $n$ one recovers the polynomials (45), which have
been correctly calculated from first principles. We fail to see the
mathematical justification of this procedure and therefore state it
simply as an ``experimental'' fact which does allow us to calculate the
averages $\langle u|(C_1)^n|u\rangle$, $n=1,\,2,\,3,\,\dots\,.$ This
calculation is much easier than the calculation which starts from first
principles.

\bigskip\medskip

I am greatly indebted to Professor Pawel O. Mazur from the Department of
Physics and Astronomy, University of South Carolina, for many useful
discussions and for having created for me excellent working conditions
at Columbia, SC, where the most important parts of this paper were
written. I am also indebted to Professor Jacek Wosiek and dr Andrzej
Rostworowski from the Department of Physics, Jagellonian University, for
many useful discussions.

\end{document}